\documentclass[aps,preprint,onecolumn,showpacs]{revtex4}
\usepackage{graphicx}
\usepackage{dcolumn}
\usepackage{amsmath}
\begin{document}

\title[Short Title]{Dynamical age of the universe as a constraint on the parametrization of dark energy equation of state\\}
%\title[Short Title]{Manuscript Title:\\
%with Forced Linebreak}% Force line breaks with \\

%\author{Ann  Author}
% \thanks{Also at Physics Department, XYZ University.}%Lines break automatically or can be forced with \\
\author{V. B. Johri}%
 \email{vinodjohri@hotmail.com}
\affiliation{%
Department of Mathematics and Astronomy, University of
Lucknow, Lucknow-226007, India\\
}%

\author{P. K. Rath}
 \email{pkrath_lu@yahoo.co.in}
\affiliation{Department of Physics, University of Lucknow, Lucknow-226007,
India\\
}%

\date{\today}

%\title{Influence of dark energy on the dynamical age of the universe\\
%&}
%\author{V. B. Johri \and P. K. Rath}
%&\date{\today}

\begin{abstract}
The dynamical age of the universe depends upon the rate of the expansion of
the universe, which explicitly involves the dark energy equation of state
parameter $w(z)$. Consequently, the evolution of $w(z)$ has a direct imprint
on the age of the universe. We have shown that the dynamical age of the
universe as derived from CMB data can be used as an authentic criterion,
being independent of the priors like the present value of the Hubble
constant $H_{0}$ and the cosmological density parameter $\Omega _{M}^{0}$, to
constrain the range of admissible values of  $w$ for quiessence models and
to test the physically viable parametrizations of the equation of state $w(z)
$ in kinessence models. An upper bound on variation of dark energy density
is derived and a relation between cosmological density parameters and the
transition redshift is established.
\end{abstract}
\maketitle

\section{Introduction}

The dark energy is now well established as a dominant constituent of the
present day universe. Its existence is directly inferred from the
acceleration in cosmic expansion as indicated by SNIa observation \cite{perl98,ries04} and indirectly from CMB anisotropy measurements \cite{sper03} and Integrated Sachs-Wolfe effect \cite{scra03}. But, the nature of dark energy still remains enigmatic whether it arises from the cosmological constant, quintessence or phantom fields with constant equation of state parameter or varying $w(z)$. The most commonly used method to explore the dark energy models in the literature is
to assume an adhoc equation of state $w(z)$$\equiv$$p_{X}/\rho _{X}$ for dark energy and parametrize $w(z)$ or dimensionless dark energy function $f(z)$$\equiv$$\rho _{X}(z)/\rho _{X}^{0}$ by one, two or more free parameters and then constrain these parameters by fitting to the observed SNIa, CMB and LSS data. Using this method Wang and Tegmark \cite{wang04} have discussed four viable parametrizations and put tight constraints over $%
f(z)$ in a rather model independent way. We have adopted a different
approach in this paper to constrain $w(z)$ using dynamical age of the
universe as the criterion to test the physical viability of different
parametrizations. First, we argue that the age of the universe known
observationally from various independent methods is found to be reasonably
consistent and it can be used as an effective constraint on the evolution of
the dark energy equation of state $w(z)$ because the dynamical age of the
universe depends on the Hubble expansion rate, which essentially involves $%
w(z)$. The possibility of using the age of the universe as a constraint has
been discussed in recent papers [3,6-12]. The oldest globular clusters
yield ages of about 12.5 Gyr with an uncertainty of 1.5 Gyr \cite{krau01}.
Assuming the genesis of dark energy from the cosmological constant ($w=-1$),
the SN Ia data yield the product of age and Hubble constant $%
H_{0}t_{0}=0.96\pm 0.04$ \cite{tonr03}. Taking $H_{0}=72$ Mpc$^{-1}$ km s$%
^{-1}$, it gives $t_{0}=13.04\pm 0.5$ Gyr. Padmanabhan \cite{padm03} has
given the maximum likelihood for $H_{0}t_{0}=0.94$ based on the analysis of
SN Ia observations. This can be consistent with $\Omega =1$ models only if $%
\Omega _{M}^{0}\simeq 0.3$ and $\Omega _{X}^{0}\simeq 0.7$. It is noteworthy
that the SNe observations constrain the combination $H_{0}t_{0}$ better than
the individual parameters. With the implicit assumption of the cosmological
constant ($w=-1$), the WMAP data \cite{free03} yields $t_{0}=13.7\pm 0.2$ Gyr.
Relaxation of this constraint would lead to variable estimates of $t_{0}$
depending upon the prior choice of $H_{0}$, $\Omega _{M}^{0}$ and $\Omega
_{X}^{0}$. On the other hand, Knox et al \cite{knox01} have given a method for precise determination of the dynamical age of the universe from CMB anisotropy
measurements. The best estimate of the dynamical age of the universe, coming
from CMB data, is $t_{0}=14.0\pm 0.5$ Gyr. Although this method also
presumes the cosmological constant, it is shown that the variation of the
equation of state away from $w=-1$ at fixed $\theta _{s}$ (the angle
subtended by the acoustic horizon on the last scattering surface) has very
little effect on the age of the universe. Moreover, the CMB method of age
determination does not involve the observational parameters $H_{0}$ and $%
\Omega _{M}^{0}$ and is therefore free from observational uncertainties
involved in the measurement of these parameters. These diverse observations
leading to mutually agreeable estimates of the age of the universe reinforce
our confidence in using the dynamical age of the universe as an effective
tool to constrain the equation of state and the dark energy density
parameter $\Omega _{X}$. We would take $t_{0}=14.0\pm 0.5$ Gyr (based on CMB
anisotropy observations \cite{knox01}) as the most reliable estimate of the
present age of the universe since it is not sensitive to variation in the
values of the Hubble constant $H_{0}$ and the fractional energy density $%
\Omega _{M}^{0}$ of the non-relativistic matter in the universe . We would
use it as the standard criterion for the dynamical age of the universe for
comparison with a variety of kinessence models \cite{sahn03} whose equation
of state $w(z)$ is assumed to have different parametrizations as discussed
in Sec. IV of the paper.

Although, the present dark energy density $\rho _{X}^{0}=\left( 4.8\pm
1.2\right) \times 10^{-30}$ $gm/cm^{3}$ is precisely known \cite{wang04} and
the cosmological density parameter $\Omega _{X}^{0}=0.73$ from WMAP data 
\cite{free03}, we have no definite estimate of $\rho _{X}(z)$ and $\Omega
_{X}(z)$ at small $z$ or large $z$ in the past. As such, we do not know
whether the dark energy evolves with cosmic time or not and if it does then
what is the mode of its evolution. However, it is possible to find $\Omega
_{X}(z)$ precisely in terms of $w(z)$ at the point of transition $z_{T}$
from the decelerating phase to accelerating expansion phase by taking $q=0$
in Eq. (8). We have discussed in our previous paper \cite{johr04}, how the
transition red shift $z_{T}$ may be used to compute the age of the universe $%
t_{m}$ up to the transition epoch and also the dynamical age of the universe 
$t_{0}$ up to the present epoch. We have used this technique in Sec. III of
this paper to investigate the range of admissible values of $w$ for
quiessence model, which are compatible with the range of dynamical age of
the universe. The same constraint is applied in Sec. IV to test the
physically viable parametrizations of $w(z)$ assuming the slow roll down
condition for the scalar field in kinessence models.

\section{Expansion dynamics of the dark energy}

During the matter dominated era onwards, the contribution of CMB photons and
neutrinos to the cosmic energy density is trivial and the major energy
constituents are non-relativistic matter (baryonic matter and dark matter)
and the dark energy. The Friedmann equations for a spatially flat ($k=0$)
universe are 
\begin{eqnarray}
H^{2}\;\; &=&\frac{8\pi G}{3}\;\left[ \rho _{M}+\rho _{X}\right]   \nonumber
\\
&=&\;\;H_{o}^{2}\left[ \Omega _{M}^{o}(1+z)^{3}+\Omega _{X}^{o}f(z)\right] 
\end{eqnarray}
and 
\begin{eqnarray}
\frac{\stackrel{..}{a}}{a} &=&-\frac{4\pi G}{3}\;\left[ \rho _{M}+\rho
_{X}\left( 1+3w\right) \right]   \nonumber \\
&=&-\frac{H^{2}}{2}\left[ \Omega _{M}+\Omega _{X}\left( 1+3w\right) \right] 
\end{eqnarray}
where $H(z)$ is the Hubble parameter. The energy density of the
non-relativistic matter $\rho _{M}$ is given by 
\begin{equation}
\rho _{M}(z)=\rho _{M}^{0}(1+z)^{3}
\end{equation}
and the dark energy density $\rho _{X}$ is written as 
\[
\rho _{X}(z)=\rho _{X}^{o}f(z)
\]
with 
\begin{equation}
f(z)\;\;=\;\;\exp \left[ 3\int_{0}^{z}{\frac{1+w(z^{\prime })}{1+z^{\prime }}%
dz^{\prime }}\right]   \label{dade}
\end{equation}
In particular for quiessence models ($w=constant$) 
\begin{equation}
f(z)\;\;=\;\;\left( 1+z\right) ^{3(1+w)}
\end{equation}
This condition holds good during `tracking' wherein slow varying equation of
state ($w\simeq cons\tan t$) is a pre-requisite.

In case of the cosmological constant ($w=-1$), $f(z)=1$ and $\rho _{X}=\rho
_{\Lambda }=cons\tan t$. In all other cases, the dark energy density $\rho
_{X}$ evolves with the redshift both for varying and non-varying $w(z)$.
Logarithmic differentiation of Eq. (4) yields 
\begin{equation}
w(z)=-1+\frac{1+z}{3}\frac{1}{\rho _{X}}\frac{d\rho _{X}}{dz}
\end{equation}
Using Eq. (6), a suitable parametrization for the dark energy density can be
assumed to find the $w(z)$ which conforms to observational constraints.
Using the cosmic energy density parametric relation $\Omega _{M}+\Omega
_{X}=1,$ $\frac{\rho _{M}}{\rho _{X}}=\frac{\Omega _{M}}{\Omega _{X}}$ and
for a spatially flat universe, Eq. (6) may be written in the form 
\begin{equation}
w(z)=\frac{1+z}{3}\frac{\Omega _{X}^{\prime }}{\Omega _{X}\left( 1-\Omega
_{X}\right) }
\end{equation}
where 
\[
\Omega _{X}^{\prime }=\frac{d\Omega _{X}}{dz}
\]
It is a significant relation as it reveals how the dark energy density
varies with the evolution of the equation of state parameter $w(z)$. With
the help of Eq. (7), the Friedmann Eq. (2) can be recast in the form 
\begin{eqnarray}
2q-1\;\; &=&\;3\;w(z)\Omega _{X}  \nonumber \\
&=&-\frac{d\ln \Omega _{M}}{d\ln (1+z)}
\end{eqnarray}

Finally, the dynamical age of the universe is given by 
\begin{eqnarray}
t_{0}\;\; &=&\;\;\int_{0}^{\infty }\frac{dz}{(1+z)H(z)}  \nonumber \\
&=&H_{0}^{-1}\;\int_{0}^{\infty }\frac{dz}{(1+z)\left[ \Omega
_{M}^{o}(1+z)^{3}+\Omega _{X}^{o}f(z)\right] ^{1/2}}  \label{age}
\end{eqnarray}
The impact of the evolution of dark energy on the dynamical age of the
universe can be clearly seen from the fact that for a given functional form
of $f(z)$, the contribution of $\rho _{X}$ to $H(z)$ in Eq. (1) goes on
decreasing with more negative values of $w(z)$. Consequently, the expansion
age of the universe given by Eq.(9) increases with evolution of $w(z)$ to
more negative values. At the same time, the cosmic expansion accelerates
according to Eq. (2) for more negative values of $w(z)$. Thus, the expansion
dynamics of the universe \cite{cope06} revolves around the equation of state
parameter $w(z)$.

\section{Quiessence Models}

Quiessence models \cite{sahn03} of dark energy ($w=constant$) find wide
application in tracker field theory \cite{sahn99,stei00,johr02} wherein slow
roll down condition of scalar fields demands $w\simeq constant$. Melchiorri
et al. \cite{melc03} have combined constraints from CMB observations
(including latest WMAP data), large scale structure, luminosity measurement
of SN type Ia and Hubble Space Telescope measurements to find the bounds of
dark energy equation of state parameter to be $-1.38<w<-0.82.$ On the basis
of WMAP data for the dynamical age of the universe, Johri \cite{johr04} has
shown that $w$ lies in a thin strip around $w=-1$, viz

\begin{equation}
-1.18<w<-0.93
\end{equation}
taking $H_{0}^{-1}=13.65$ Gyr.

At the transition epoch ($q=0$), Eq. (8) simplifies to the differential
equation 
\begin{equation}
\frac{dw}{dx}=-\frac{w(1+3w)}{1+x}
\end{equation}
where $x=z_{T}$ is the redshift at the transition from decceleration to
accelerating expansion phase corresponding to equation of state parameter $w$%
. A particular integral of Eq. (11) yields

\begin{equation}
1+z_{T}=\left[ -(1+3w)\frac{\Omega _{X}^{0}}{\Omega _{M}^{0}}\right] ^{-1/3w}
\end{equation}
In fact, Eq. (12) follows directly from Eq. (2) for quiessence models. For a
prior choice of the ratio $\frac{\Omega _{X}^{0}}{\Omega _{M}^{0}}$, it is
interesting to plot the variation of $z_{T}$ versus $w$ according to the Eq.
(12) as shown in our previous paper \cite{johr04}. Eq. (11) gives the
gradient of the $z_{T}$ $\sim $ $w$ curve. This equation can be
alternatively written as 
\begin{equation}
\frac{1}{\Omega _{M}(z_{T})}\left. \frac{d\Omega _{M}}{dz}\right| _{z=z_{T}}=%
\frac{1}{1+z_{T}}
\end{equation}
and 
\begin{equation}
\Omega _{X}(z_{T})=1+(1+z_{T})\left. \frac{d\Omega _{X}}{dz}\right|
_{z=z_{T}}
\end{equation}
In Table 1, we have given values of transition redshift $z_{T}$ for various
quiessence models with dark energy equation of state parameter $w,$ the dark
energy density parameter $\Omega _{X}(z_{T})$ and corresponding age of the
universe $t_{0}$ taking $H_{0}^{-1}=13.77$ Gyr. We have also plotted the
variation of $z_{T}$ and $t_{0}$ with respect to $w$ in Fig. 1. It is
noticed that there exists an inverse correlation between the variation of $%
z_{T}$ and $t_{0}$ with respect to $w$ in quiessence models of dark energy.
Jimenez et al. \cite{jime03} have detected a similar age-redshift
correlation between the ages of the oldest galaxies and their redshifts.

\section{Kinessence Models}

In case of kinessence models \cite{sahn03} with varying equation of state $%
w=w(z)$, there are various competing dark energy models with different
parametrizations of $w(z)$ such as one index parametrizations by Gong et al. 
\cite{gong05}, two index parametrizations by Huterer et al. \cite{hute00},
Weller et al. \cite{well02}, Chevallier et al. \cite{chev01}, Linder \cite
{lind03}, Jassal et al. \cite{jass04}, Upadhye et al. \cite{upad04} as well
as Wetterich \cite{wett04} and four index parametrizations by Hannestad et
al. \cite{hann04} and Lee \cite{lee05}. The question is how well the dark
energy equation of state and the cosmological density parameter $\Omega _{X}$
in these models behave with increasing redshift $z$. We have discussed the
critical evaluation of some of these models in our earlier work \cite{johr05}%
. In the following section, we examine the cosmic age predictions of these
models given in Table 2 and compare them with the observational estimates of
the dynamical age of the universe $t_{0}=14.0\pm 0.5$ Gyr derived from CMB
anisotropy measurements \cite{knox01}.

\subsection{One index parametrizations}

\textbf{1. Gong-Zhang 1st parametrization}

The one index dark energy equation of state parameter $w(z)$ is given by
Gong et al \cite{gong05} as 
\begin{equation}
w(z)=\frac{w_{0}}{1+z}  \label{gong1}
\end{equation}
The best fit values to SN Ia `gold set', SDSS and WMAP data are $w_{0}=-1.1$
and $\Omega _{M}^{o}=0.25$. Hence, the parameters favor dark energy of
phantom origin. Combining Eqs. (\ref{gong1}) and (\ref{dade}), the dark
energy density is given by 
\begin{equation}
\rho _{X}\left( z\right) =\rho _{X}^{0}\left( 1+z\right) ^{3}\exp \left[ 
\frac{3w_{0}z}{1+z}\right]
\end{equation}
and the dark energy parameter $\Omega _{X}\left( z\right) $ can be written
as 
\begin{equation}
\Omega _{X}\left( z\right) =\left[ 1+\frac{\Omega _{M}^{0}}{\Omega _{X}^{0}}%
\exp (\frac{-3w_{0}z}{1+z})\right] ^{-1}
\end{equation}
We calculate the transition redshift $z_{T}=0.56$ for this model by taking $%
q=0$ in Eq.(8) and inserting for $w$ and $(\Omega _{X})_{T}$ (the best fit
values of the parameters $w_{0}$ and $\Omega _{M}^{0}$ ) from Eqs. (15) and
(17) respectively. Knowing the transition redshift $z_{T}$, the total
dynamical age of the universe ( worked out using the technique of \cite
{johr04}) turns out to be $13.33$ Gyr. Since it lies outside the
observational range given by Knox et al \cite{knox01}, this parametrization
is found to be incompatible with the age constraint. Following the same
procedure, we have found the dynamical age of the universe for various
parametrizations discussed in this section and have tested their physical
viability on the basis of the age constraint. The results are summarized in
Table 2.
 
\textbf{2. Gong-Zhang 2nd parametrization}

The second one index dark energy equation of state parameter $w(z)$ is given
by \cite{gong05} 
\begin{equation}
w\left( z\right) =\frac{w_{0}}{1+z}\exp \left( \frac{z}{1+z}\right)
\label{gong2}
\end{equation}
The best fit values to SN Ia `gold set', SDSS and WMAP data are $w_{0}=-0.97$
and $\Omega _{M}^{o}=0.28$ and the parameters favor dark energy of
quintessence origin. Inserting for $w(z)$ from Eq. (\ref{gong2}) into Eq. (%
\ref{dade}), one gets \qquad 
\begin{equation}
\rho _{X}=\rho _{X}^{0}\left( 1+z\right) ^{3}e^{3w_{o}\frac{z}{1+z}-3w_{0}}
\end{equation}
and the dark energy parameter $\Omega _{X}\left( z\right) $ turns out to be

\begin{equation}
\Omega _{X}\left( z\right) =\left[ 1+\frac{\Omega _{M}^{0}}{\Omega _{X}^{0}}%
\exp \left( 3w_{0}-3w_{0}e^{\frac{z}{1+z}}\right) \right] ^{-1}
\end{equation}
For this parametrization, the transition red shift $z_{T}=0.39$, we get $%
(\Omega _{X})_{T}=0.383$. the age of the universe $t_{0}=13.30$ Gyr.

\subsection{Two index parametrizations}

\textbf{1. Linear-redshift parametrization}

The dark energy equation of state parameter $w(z)$ is given by Huterer et
al. \cite{hute00} and Weller et al. \cite{well02} as 
\begin{equation}
w(z)\;\;=\;\;w_{o}+w^{\prime }z,\quad \quad w^{\prime }=\Big( \frac{dw}{dz}%
\Big)_{z=0}  \label{hute}
\end{equation}
It has been used by Riess et al. \cite{ries04} for probing SN Ia
observations at $z<1$. The best fit values to SN Ia `gold set' data \cite
{dicu04} are $w_{o}=-1.40$, $w^{\prime }=1.67$ and $\Omega _{M}^{o}=0.30$.
Hence, the parameters favor dark energy of phantom origin. Inserting Eqs. (%
\ref{hute}) into Eq.(\ref{dade}), one gets 
\begin{equation}
\rho _{X}(z)=\rho _{X}^{0}(1+z)^{3(1+w_{0}-w^{\prime })}\exp (3w^{\prime }z)
\end{equation}
and 
\begin{equation}
\Omega _{X}(z)=\left[ 1+\frac{\Omega _{M}^{0}}{\Omega _{X}^{0}}%
(1+z)^{-3(w_{0}-w^{\prime })}\exp (-3w^{\prime }z)\right] ^{-1}
\end{equation}
In this case, the transition red shift $z_{T}=0.39$, the corresponding
dynamical age of the universe is $10.98$ Gyr which lies beyond. the
observational estimates.

\textbf{2. Chevallier-Polarski-Linder parametrization}

The dark energy equation of state parameter $w(z)$ is given as \cite
{chev01,lind03} 
\begin{equation}
w(z)\;\;=\;\;w_{o}+\frac{w_{1}z}{1+z}  \label{pola}
\end{equation}
The best fit values of $w_{o},$ $w_{1}$ and $\Omega _{m}^{0}$ to SN Ia `gold
set' data \cite{gong04a,dicu04} are $-1.6$, $3.3$ and $0.30$ respectively
and the parameters suggest that the dark energy is of phantom origin. The
dark energy density is given by 
\begin{equation}
\rho _{X}(z)=\rho _{X}^{0}(1+z)^{3(1+w_{0}+w_{1})}\exp (-\frac{3w_{1}z}{1+z})
\end{equation}
and the dark energy parameter 
\begin{equation}
\Omega _{X}(z)=\left[ 1+\frac{\Omega _{M}^{0}}{\Omega _{X}^{0}}%
(1+z)^{-3(w_{0}+w_{1})}\exp (\frac{3w_{1}z}{1+z})\right] ^{-1}
\end{equation}
The transition red shift $z_{T}=0.35$ and the calculated dynamical age of
the universe $t_{0}=11.23$ Gyr lies beyond the observational estimates.

\textbf{3. Jassal-Bagla-Padmanabhan parametrization}

Jassal \textit{et al.} have parametrized $w(z)$ as \cite{jass04}

\begin{equation}
w(z)\;\;=\;\;w_{o}+\frac{w_{1}z}{(1+z)^{2}}  \label{jasl}
\end{equation}
The best fit values to SN Ia `gold set' data are $w_{o}=-1.9$, $w_{1}=6.6$
and $\Omega _{M}^{o}=0.30$. One has $w(z)=w_{0}=-1.9$ at $z=0$. The
parameters suggest that the dark energy is of phantom origin. Combining Eq. (%
\ref{jasl}) and Eq. (\ref{dade}) one obtains 
\begin{equation}
\rho _{X}(z)=\rho _{X}^{0}(1+z)^{3(1+w_{0})}\exp \left[ \frac{3w_{1}z^{2}}{%
2(1+z)^{2}}\right]
\end{equation}
and 
\begin{equation}
\Omega _{X}(z)=\left[ 1+\frac{\Omega _{M}^{0}}{\Omega _{X}^{0}}%
(1+z)^{-3w_{0}}\exp \left\{ -\frac{3w_{1}z^{2}}{2(1+z)^{2}}\right\} \right]
^{-1}
\end{equation}
The transition redshift $z_{T}=0.3$ and the corresponding age of the
universe turns out to be $12.94$ Gyr for the parametrization of Jassal et al.%
\cite{jass04}.

\textbf{4. Upadhye-Ishak-Steinhardt parametrization}

Upadhye et al. have parametrized $w(z)$ as \cite{upad04} 
\begin{equation}
w(z)=\left\{ 
\begin{array}{lll}
w_{0}+w_{1}z & if & z<1 \\ 
w_{0}+w_{1} & if & z\geq 1
\end{array}
\right.  \label{upad}
\end{equation}
The best fit values of parameters for SN Ia `gold set', galaxy power
spectrum and CMB power spectrum data are $w_{0}=-1.38$, $w_{1}=1.2$ and $%
\Omega _{M}^{0}=0.31$. Hence, the parameters suggest that the dark energy
has phantom origin. The dark energy density is given by 
\begin{eqnarray}
\rho _{x}(z) &=&\rho _{X}^{0}(1+z)^{3(1+w_{0}-w_{1})}\exp
(3w_{1}z)\,\,\,\,\,\,\,\,\,\,\,\,\, 
\begin{array}{lll}
if &  & z<1
\end{array}
\nonumber \\
&=&\rho _{X}^{0}(1+z)^{3(1+w_{0}+w_{1})}\qquad
\,\,\,\,\,\,\,\,\,\,\,\,\,\,\,\,\,\,\,\,\,\,\,\,\,\,\,\,\,\,\,\,\,\, 
\begin{array}{lll}
if &  & z\geq 1
\end{array}
\end{eqnarray}
and the dark energy density parameter is written as 
\begin{equation}
\Omega _{X}(z)=\left\{ 
\begin{array}{c}
\begin{array}{lllll}
\left[ 1+\frac{\Omega _{M}^{0}}{\Omega _{X}^{0}}(1+z)^{-3(w_{0}-w_{1})}\exp
(-3w_{1}z)\right] ^{-1} &  & if &  & z<1
\end{array}
\\ 
\begin{array}{lllll}
\left[ 1+\frac{\Omega _{M}^{0}}{\Omega _{X}^{0}}%
(1+z)^{-3(w_{0}+w_{1})}e^{-3w_{1}\left( 1-2\ln 2\right) }\right] ^{-1} &  & 
if &  & z\geq 1
\end{array}
\end{array}
\right.
\end{equation}
The transition red-shift $z_{T}$ turns out to be $0.44$ and corresponding
dynamical age of the universe turns out to be $12.87$ Gyr.

\textbf{5.} \textbf{Wetterich Parametrization}

Wetterich has parametrized the dark energy equation of state $w(z)$ as \cite
{wett04} 
\begin{equation}
w\left( z\right) =\frac{w_{0}}{\left[ 1+b\ln \left( 1+z\right) \right] ^{2}}
\label{wett}
\end{equation}
The best fit values to SN Ia `gold set' data are $w_{0}=-2.5$, $b=4.0$ and $%
\Omega _{M}^{o}=0.3$. One has $w(z)=w_{0}=-2.5$ at $z=0$. The parameters
suggest that the dark energy is of phantom origin. Inserting Eq.(\ref{wett}
) for $w(z)$ in Eq. (\ref{dade}) , one gets 
\begin{equation}
\rho _{X}(z)=\rho _{X}^{0}\left( 1+z\right) ^{3+\frac{3w_{0}}{1+b\ln \left(
1+z\right) }}
\end{equation}
and 
\begin{equation}
\Omega _{X}(z)=\left[ 1+\frac{\Omega _{M}^{0}}{\Omega _{X}^{0}}(1+z)^{-\frac{%
3w_{0}}{1+b\ln \left( 1+z\right) }}\right] ^{-1}
\end{equation}
For $z_{T}=0.26$, the calculated age of the universe $t_{0}=12.52$ Gyr in
comparison to the age $14.0\pm 0.5$ Gyr derived from CMB anisotropies by
Knox et al \cite{knox01}.

\subsection{Four index parametrizations}

\textbf{1. Hannestad-M\"{o}rtsell parametrization}

Let us now consider Hannestad parametrization \cite{hann04} which involves
4-parameters. %%%%%%%%%%%%%%%%%%%%%%%%%%%%%%%%%%%%%%%%%%
\begin{eqnarray}
w(z)\;\; &=&\;\;w_{o}w_{1}\frac{a^{p}+a_{s}^{p}}{w_{1}a^{p}+w_{o}a_{s}^{p}}\;
\nonumber \\
\; &=&\;\;\frac{1+\big(\frac{1+z}{1+z_{s}}\big)^{p}}{w_{o}^{-1}+w_{1}^{-1}%
\big(\frac{1+z}{1+z_{s}}\big)^{p}}
\end{eqnarray}
where $w_{0}$ and $w_{1}$ are the asymptotic values of $w(z)$ in the distant
future ($1+z\rightarrow 0$) and in the distant past ($z\rightarrow \infty $)
respectively. The $a_{s}$ and $p$ are the scale factor at the change over
and the duration of the change over in $w$ respectively. Taking the best fit
values for the combined CMB, LSS and SN Ia data \cite{hann04}, $%
w_{o}=\;-1.8, $ $w_{1}=\;-0.4$ $,$ $q\;=\;3.41$ and $a_{s}=0.50$ with a
prior $\Omega _{M}^{o}=0.38$, one has $w(z=0)=-1.38$ at $z=0.$ These
parameters suggest that the dark energy is of phantom origin. Further, the
dark energy density $\rho _{x}(z)$ is given by 
\begin{equation}
\rho _{X}(z)=\rho _{X}^{0}(1+z)^{3(1+w_{1})}\times \left[ \frac{\left(
w_{1}+w_{0}a_{s}^{p}\right) \left( 1+z\right) ^{p}}{w_{1}+w_{0}a_{s}^{p}%
\left( 1+z\right) ^{p}}\right] ^{\frac{3\left( w_{0-}w_{1}\right) }{p}}
\end{equation}
and the dark energy parameter is written as 
\begin{equation}
\Omega _{X}(z)=\left[ 1+\frac{\Omega _{M}^{0}}{\Omega _{X}^{0}}%
(1+z)^{-3w_{1}}\left\{ \frac{w_{1}+w_{0}a_{s}^{p}\left( 1+z\right) ^{p}}{%
\left( w_{1}+w_{0}a_{s}^{p}\right) \left( 1+z\right) ^{p}}\right\} ^{\frac{%
-3\left( w_{0-}w_{1}\right) }{p}}\right] ^{-1}
\end{equation}
At transition redshift $z_{T}=0.39$, the dark energy parameter $(\Omega
_{X})_{T}=0.333$. The age of the universe $t_{0}=12.52$ Gyr in comparison to 
$t_{0}=14.0\pm 0.5$ Gyr derived by Knox et al \cite{knox01} from CMB
observations.

\textbf{2. Lee parametrization}

The dark energy equation of state parameter $w(z)$ is parametrized by Lee as 
\cite{lee05} 
\begin{equation}
w(z)=w_{r}\frac{w_{0}\exp (px)+\exp (px_{c})}{\exp (px)+\exp (px_{c})}
\label{lee}
\end{equation}
where 
\begin{equation}
x=\ln a=-\ln (1+z)
\end{equation}
and the symbols $w_{r}$ and $x_{c}$ have the usual meaning as in \cite{lee05}%
. The parameters $w_{r}$ is chosen to be $1/3$ using the tracking condition
and $w_{0}$ is taken as $-3$. The other parameters are obtained by analyzing
the separation of CMB\ peaks and the time variation of the fine structure
constant. For $\Omega _{M}^{0}=0.27$ and $x_{c}=-2.64$, the range of $p$ is
taken as $1.5\leq p\leq 3.9$. At $z=0$, one has $w(z=0)=-0.832$. Hence, the
parameters suggest that the dark energy is of quintessence origin. Inserting
Eq. (\ref{lee}) into Eq. (\ref{dade}) , one gets 
\begin{equation}
\rho _{X}\left( z\right) =\rho _{X}^{0}\left( 1+z\right)
^{3}(a+a_{eq})\left( \frac{a^{p}+a_{c}^{p}}{1+a_{c}^{p}}\right) ^{\frac{4}{p}%
}
\end{equation}
The dark energy parameter $\Omega _{X}(z)$ is given by 
\begin{equation}
\Omega _{X}(z)=\left[ 1+\frac{\Omega _{M}^{0}}{\Omega _{X}^{0}}%
(a+a_{eq})\left( \frac{a^{p}+a_{c}^{p}}{1+a_{c}^{p}}\right) ^{-4/p}\right]
^{-1}
\end{equation}
For $p=1.5$ and $3.9$, the dark energy parameter $\Omega _{X}$ is 0.353 and
0.333 at the transition red shift $z_{T}=0.74-0.76$ respectively. In
comparison to the observational estimate of the age derived by Knox et al 
\cite{knox01} $t_{0}=14.0\pm 0.5$ Gyr, the age of the universe for this
parametrization turns out to be $13.57$ and $13.67$ Gyr for $p=1.5$ and $3.9$
respectively.

\section{Bounds on variation of dark energy}

In the case of kinessence models with equation of state $w=w(z)$, $\rho _{X}$
and $\Omega _{X}$ are expressible in terms of $f(z)$ which depends upon the
choice of parametrization of $w(z)$. So, it is not possible to predict the
variation of $\Omega _{X}$ $(z)$ precisely in the past without knowing the
realistic form of $f(z)$. Various parametrizations as discussed in the Sec.
4.1 and 4.2 give only the tentative variation of $\rho _{X}$ and $\Omega
_{X} $ with the redshift. However, we can put an upper bound on the
variation of $\Omega _{X}(z)$ in the quintessence models. According to the
Table 1, the dynamical age of the universe given by Knox et al. \cite{knox01}
constrains $\ w(z)$ to lie within the range $-1.60\leq w(z)\leq -0.82$ which
is consistent with WMAP upper bound $w<-0.8$ \cite{free03} and satisfies the
dark energy condition $w(z)<-2/3$. Applying this condition to the integral
in Eq. (4) leads to an upper bound on the dimensionless dark energy function $f(z)$ given by 
\begin{equation}
f(z)<(1+z)
\end{equation}
Combining $\rho _{X}(z)\leq \rho _{X}^{0}\left( 1+z\right) $ with $\rho
_{M}(z)=\rho _{M}^{0}\left( 1+z\right) ^{3}$ we get the upper bound on $\rho
_{X}$ and $\Omega _{X}$%
\begin{equation}
\frac{\rho _{X}}{\rho _{M}}\leq \frac{\rho _{X}^{0}}{\rho _{M}^{0}}(1+z)^{-2}
\end{equation}
and 
\begin{equation}
\Omega _{X}^{-1}\geq 1+\frac{\Omega _{M}^{0}}{\Omega _{X}^{0}}(1+z)^{2}
\end{equation}

As shown in the Fig. 2, to check the validity of Eq. (45), let us choose a
quintessence model with $w=-0.93=$constant. The corresponding $\Omega
_{X}=0.358$ from Table 1 at the transition redshift $z_{T}=0.760$ whereas
Eq. (45) yields the upper bound $\Omega _{X}\leq 0.46$ for $z_{T}=0.760$, $%
\Omega _{M}^{0}=0.27$ and $\Omega _{X}^{0}=0.73$. Eq. (45) can also be
applied, in principle, to dark energy models with varying equation of state
parameter $w(z)$ using well-behaved parametrizations to yield the upper
bound of the fractional dark energy density $\Omega _{X}$.

\section{Conclusions}

A new approach to the exploration of dark energy parameters is
discussed based on the impact of the evolution of dark energy equation of
state $w(z)$ on the dynamical age of the universe. The dynamical age depends
upon the Hubble expansion of the universe of which dark energy is presently
a dominant constituent. Therefore, the age of the universe carries the
signature of the dark energy and can be used as an effective constraint to
check the physical viability of the various quiessence and kinessence models
as discussed in this paper. Unlike the conventional procedure of assuming a
parametrization for $w(z)$ and constraining it by fitting it to the
observational data, we have calculated the dynamical age of the universe
assuming different parametrizations and tested the physical viability of
these parametrizations with the cosmic age constraint derived from the CMB
data \cite{knox01}. Direct comparison of the theoretically calculated age of the
universe (Table 1 and 2) with the observational range of the cosmic age as
laid down by CMB anisotropy measurements reveals that Lee parametrizations
\cite{lee05} satisfy both the astrophysical \cite{johr05} and the cosmic age constraint \cite{knox01}. Of course, the quintessence models in the range $-1<w<-0.82$, the cosmological constant and a wide class of phantom models also satisfy the cosmic age constraint.

Recently Jassal, Bagla and Padmanabhan \cite{jass05} have carried out a
detailed analysis of constraints on cosmological parameters from different
observations with particular reference to equation of state parameter $w(z)$%
. According to their analysis, the SN Ia observations alone favor phantom
models ($w\ll -1$) with large $\Omega _{M}$ whereas the WMAP observations
favor models with $w\sim -1$ if dark energy perturbations are included. In
the absence of any definite evidence of variation of dark energy density
with the redshift, the cosmological constant still remains a favorite
candidate for dark energy \cite{padm05} as it remains compatible with
astrophysical observations \cite{johr05} and satisfies the cosmic age
constraint as laid down in this paper as well as the recently published WMAP
three years data \cite{sper06} estimating $t_{o}=13.73_{-0.17}^{+0.13}$ $Gyr$
and $H_{0}=73.4_{-3.8}^{+2.8}$ Mpc$^{-1}$ km s$^{-1}$.

\begin{acknowledgments}
The authors gratefully acknowledge the facilities and financial assistance
for carrying out the research work done under the DST project at University of Lucknow, Lucknow, India. The authors also thank the referee for the
constructive suggestions for the improvement of paper.
\end{acknowledgments}

\textbf{Table 1:} Age of the universe with constant dark energy parameter $w$
and $H_{0}^{-1}=13.77$ Gyr.

\ 
\begin{tabular}{cccccc}
\hline\hline
$w$ & $z_{T}$ & $(\Omega _{M})_{T}$ & $t_{m}$ & $t_{0}H_{0}$ & $t_{0}$ (Gyr)
\\ \hline\hline
&  &  &  &  &  \\ 
-0.66 & 0.636 & 0.495 & 0.613 & 1.128 & 15.53 \\ 
-0.70 & 0.680 & 0.524 & 0.589 & 1.106 & 15.22 \\ 
-0.75 & 0.718 & 0.556 & 0.570 & 1.082 & 14.91 \\ 
-0.80 & 0.741 & 0.583 & 0.558 & 1.064 & 14.65 \\ 
-0.85 & 0.754 & 0.608 & 0.552 & 1.049 & 14.44 \\ 
-0.90 & 0.759 & 0.629 & 0.550 & 1.036 & 14.27 \\ 
-0.93 & 0.760 & 0.642 & 0.550 & 1.030 & 14.18 \\ 
-0.95 & 0.759 & 0.649 & 0.550 & 1.026 & 14.13 \\ 
-1.00 & 0.755 & 0.667 & 0.552 & 1.017 & 14.01 \\ 
-1.02 & 0.753 & 0.673 & 0.553 & 1.014 & 13.97 \\ 
-1.05 & 0.749 & 0.683 & 0.555 & 1.010 & 13.91 \\ 
-1.10 & 0.740 & 0.697 & 0.559 & 1.005 & 13.84 \\ 
-1.15 & 0.730 & 0.710 & 0.564 & 1.000 & 13.77 \\ 
-1.18 & 0.723 & 0.718 & 0.567 & 0.998 & 13.74 \\ 
-1.20 & 0.719 & 0.722 & 0.569 & 0.996 & 13.72 \\ 
-1.35 & 0.684 & 0.753 & 0.587 & 0.989 & 13.62 \\ 
-1.50 & 0.648 & 0.778 & 0.607 & 0.986 & 13.57 \\ 
-1.60 & 0.625 & 0.792 & 0.620 & 0.985 & 13.56 \\ 
-1.70 & 0.603 & 0.804 & 0.632 & 0.985 & 13.57 \\ 
-1.80 & 0.582 & 0.815 & 0.645 & 0.986 & 13.58 \\ 
-1.90 & 0.562 & 0.825 & 0.657 & 0.987 & 13.60 \\ 
-2.00 & 0.543 & 0.833 & 0.669 & 0.989 & 13.62 \\ 
-2.10 & 0.526 & 0.841 & 0.681 & 0.991 & 13.65 \\ 
-2.20 & 0.509 & 0.849 & 0.692 & 0.994 & 13.68 \\ 
-2.30 & 0.494 & 0.855 & 0.703 & 0.996 & 13.72 \\ 
-2.40 & 0.479 & 0.861 & 0.713 & 0.999 & 13.75 \\ 
-2.50 & 0.465 & 0.867 & 0.723 & 1.001 & 13.79 \\ \hline\hline
\end{tabular}

%\pagebreak
\vspace{10 mm} \noindent \textbf{Table 2:} The age of the universe in
parametric models of dark energy.\textbf{\ }

\begin{tabular}{cccccc}
\hline\hline
Models & Reference & $z_{T}$ & $(\Omega _{x})_{T}$ & $t_{0}H_{0}$ & $t_{0}$ $%
(Gyr)$ \\ \hline\hline
&  &  &  &  &  \\ 
\multicolumn{1}{l}{Gong-Zhang 1st} & \multicolumn{1}{l}{\cite{gong05}} & 0.56
& 0.479 & 0.968 & 13.33 \\ 
\multicolumn{1}{l}{Gong-Zhang 2nd} & \multicolumn{1}{l}{\cite{gong05}} & 0.39
& 0.383 & 0.966 & 13.30 \\ 
\multicolumn{1}{l}{Linear red shift} & \multicolumn{1}{l}{\cite
{hute00,well02}} & 0.39 & 0.443 & 0.797 & 10.98 \\ 
\multicolumn{1}{l}{Chevallier-Polarski-Linder} & \multicolumn{1}{l}{\cite
{chev01,lind03}} & 0.35 & 0.451 & 0.816 & 11.23 \\ 
\multicolumn{1}{l}{Jassal-Bagala-Padmanabhan} & \multicolumn{1}{l}{\cite
{jass04}} & 0.30 & 0.467 & 0.939 & 12.94 \\ 
\multicolumn{1}{l}{Upadhye-Ishak-Steinhardt} & \multicolumn{1}{l}{\cite
{upad04}} & 0.44 & 0.392 & 0.934 & 12.87 \\ 
\multicolumn{1}{l}{Wetterich} & \multicolumn{1}{l}{\cite{wett04}} & 0.26 & 
0.488 & 0.909 & 12.52 \\ 
\multicolumn{1}{l}{Hannestad-M\"{o}rtsell} & \multicolumn{1}{l}{\cite{hann04}
} & 0.39 & 0.333 & 0.909 & 12.52 \\ 
\multicolumn{1}{l}{Lee ($p=1.5$)} & \multicolumn{1}{l}{\cite{lee05}} & 0.74
& 0.353 & 0.985 & 13.57 \\ 
\multicolumn{1}{l}{Lee ($p=3.9$)} & \multicolumn{1}{l}{\cite{lee05}} & 0.76
& 0.333 & 0.993 & 13.67 \\ \hline\hline
\end{tabular}

\pagebreak

\begin{figure}[tbp]
\includegraphics[width=16cm]{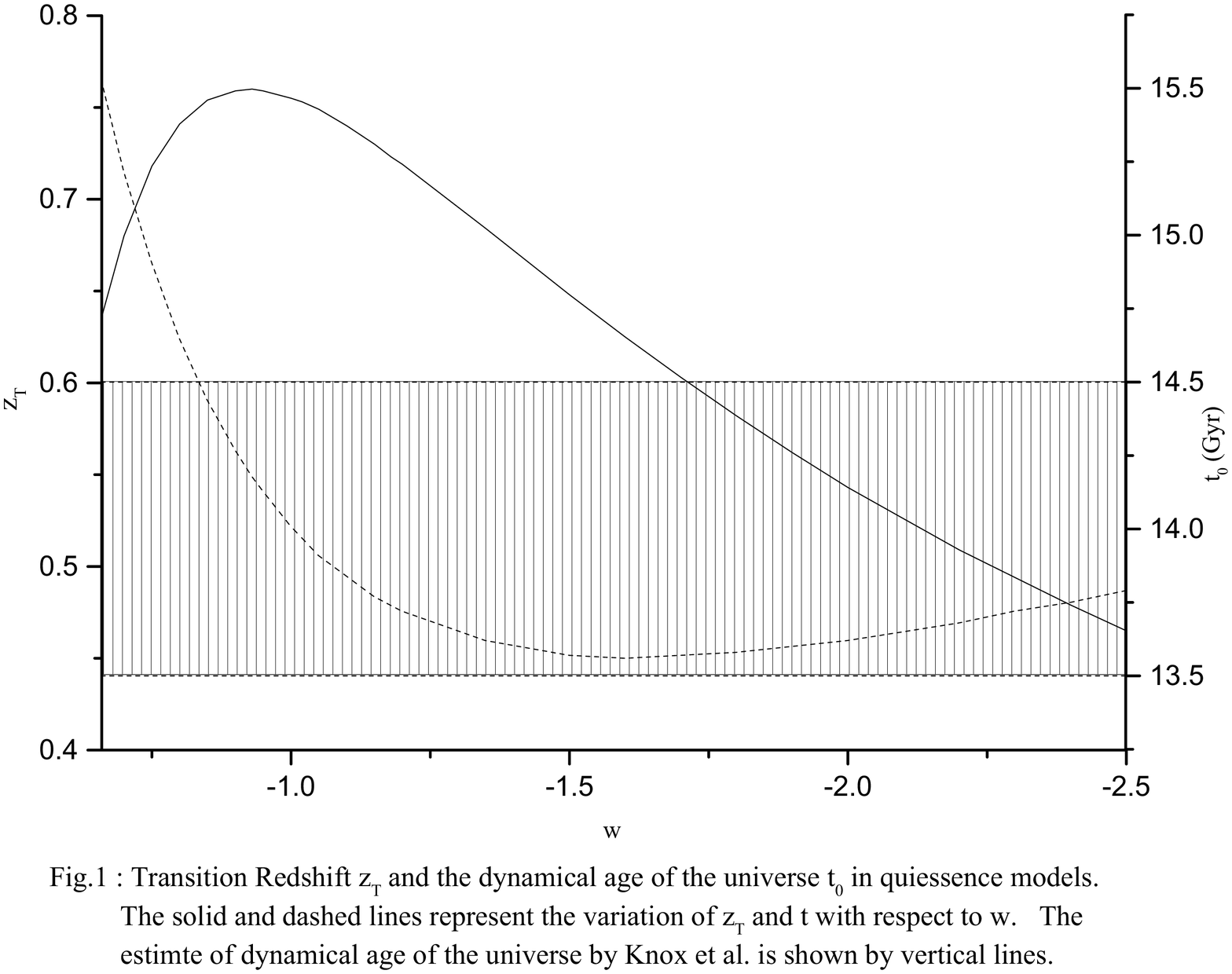} %\includegraphics{fig1.eps}
%\end{center}
\end{figure}

%\pagebreak

\begin{figure}[tbp]
\includegraphics[width=16cm]{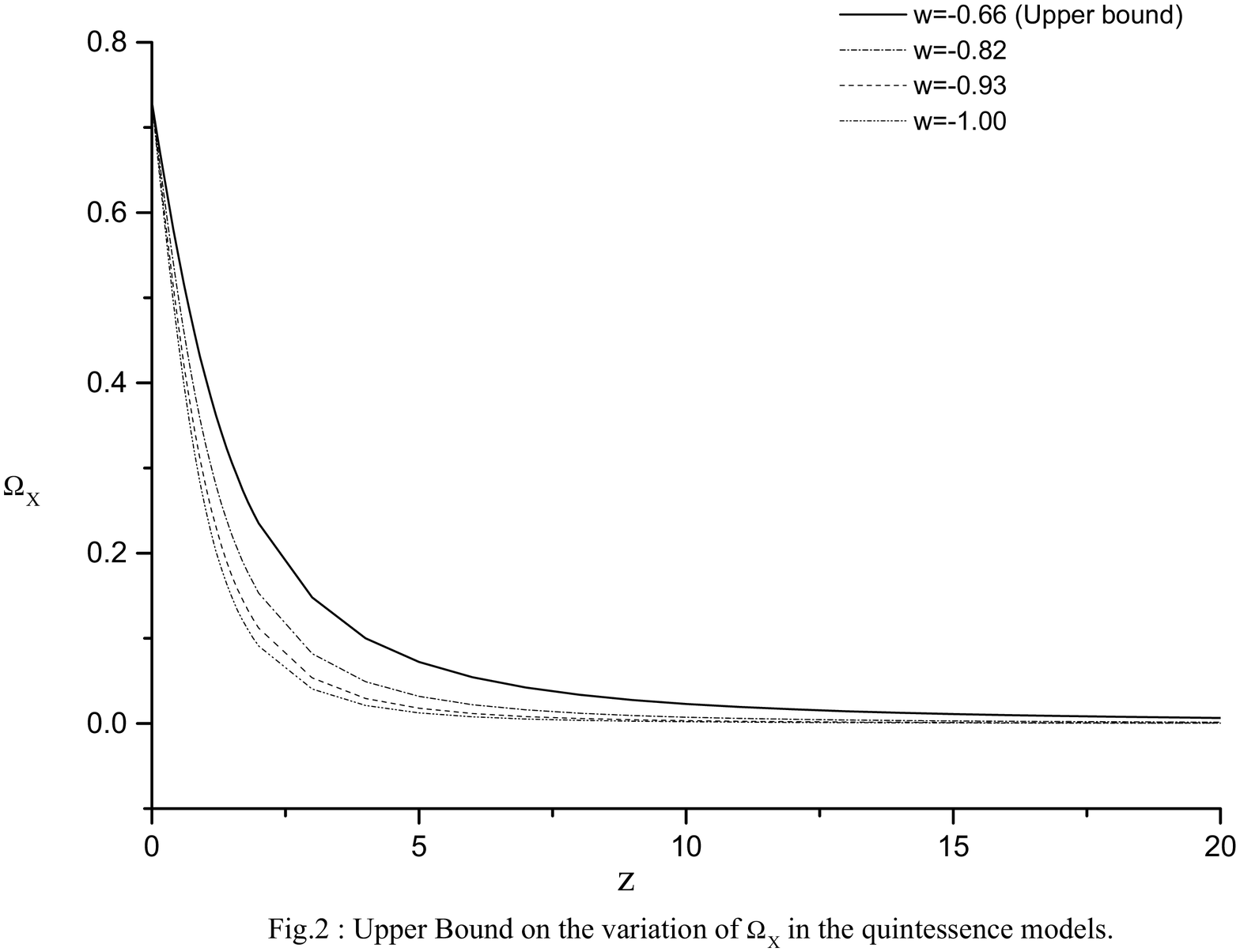} %\includegraphics{fig2.eps}
%\end{center}
\end{figure}

\end{document}